\documentclass{article}
\usepackage[margin=2.9cm]{geometry}
\usepackage[utf8]{inputenc}
\usepackage{amsmath}  
\usepackage{graphicx} 
\usepackage{siunitx}
\usepackage{subcaption}

\newcommand{\um}{\si{\micro\meter}}
\begin{document}

\title{Construction of a radiation beam scanner and investigation of volume averaging correction factor effects on beam-profile}

\author{J.~Tikkanen$^{\rm 1,2,3}$, J.~Morelius$^{\rm 1}$, T.~Siiskonen$^{\rm 1,2,3}$} 

\date{}

\maketitle

~\\
$^{\rm 1}$Radiation and Nuclear Safety Authority (STUK), Jokiniemenkuja 1, FI-01370 Vantaa, Finland \\
$^{\rm 2}$Helsinki Institute of Physics, Gustaf Hällströmin katu 2, FI-00014 University of Helsinki, Finland \\
$^{\rm 3}$University of Helsinki, Department of Physics, P.O. Box 64, FI-00014 University of Helsinki, Finland

~\\ \large{October 2022} \\

\begin{abstract}

In radiation beam-profile measurements, an accurate positioning of the detector with high position resolution is essential. For this purpose, we built a scanning device capable of moving a detector in three dimensions using mainly parts from a commercial 3D-printer. The accuracy and repeatability of movement was tested with caliper, laser displacement sensor, and repeated $^{60}$Co beam-profile measurements in a water phantom. The results from the caliper and the sensor showed position accuracy for the scanner to be better than $\pm$150~\um. The standard deviation of the error in position from laser sensor measurements was approximately 30~\um\ and the beam profile scans showed a maximum deviation from the mean position of 50~\um. The effect of volume averaging correction factors on $^{60}$Co beam-profile was investigated with two different sized ionization chambers. The differences in the profiles were reduced significantly after applying the correction factors.

\end{abstract}

%\begin{keyword}
%Dosimetry, position accuracy, dose-profile, radiotherapy, quality control
%\end{keyword}

\section{Introduction}

Measurement of dose distributions in a water phantom is an essential part in assuring that the right dose in radiotherapy is delivered to the patient. The dose distributions, or dose-profiles, are commonly measured by moving a small ionization chamber, or a semiconductor detector, in the phantom and taking measurements at multiple positions. A high position accuracy and resolution is required for these measurements \cite{TRS483}. At radiotherapy clinics, the phantoms have a built in mechanism to move the detector, and software to acquire the profiles automatically. These phantoms are usually provided by commercial manufacturers. Similar dose profile measurements are of interest in dosimetry laboratories, for example in radiation beam quality control. Since the scanning mechanism in commercial scanners is built into the phantom, these systems are not suitable for generic dosimetry in laboratory, such as air-kerma measurements free in air.

In this article, a generic beam scanner based on a commercial 3D-printer is constructed. Accurate and automated movement is essential also in 3D-printing and hence a similar device can potentially be used for beam profile measurements. Control electronics and software were self-made, utilizing stepper-motor drivers and a Raspberry Pi. 

The accuracy and repeatability of the scanner was tested with position measurements and by making repeated dose-profile measurements in a $^{60}$Co irradiator beam. Even though the scanner was built for dose-profile measurements, the device may find use in a wide range of fields requiring accurate 3D movement.

The scanner was used to compare beam profiles of the $^{60}$Co irradiator measured with two different sized ionization chambers. This was to investigate the effect of dose averaging in the chamber cavity, and to test the accuracy of the profile measured with the larger chamber after applying volume averaging correction factors. 

\begin{figure}
    \centering
    \begin{subfigure}[t]{0.45\textwidth}
        \includegraphics[ width=\textwidth, trim={20cm 0 20cm 10cm}, clip]{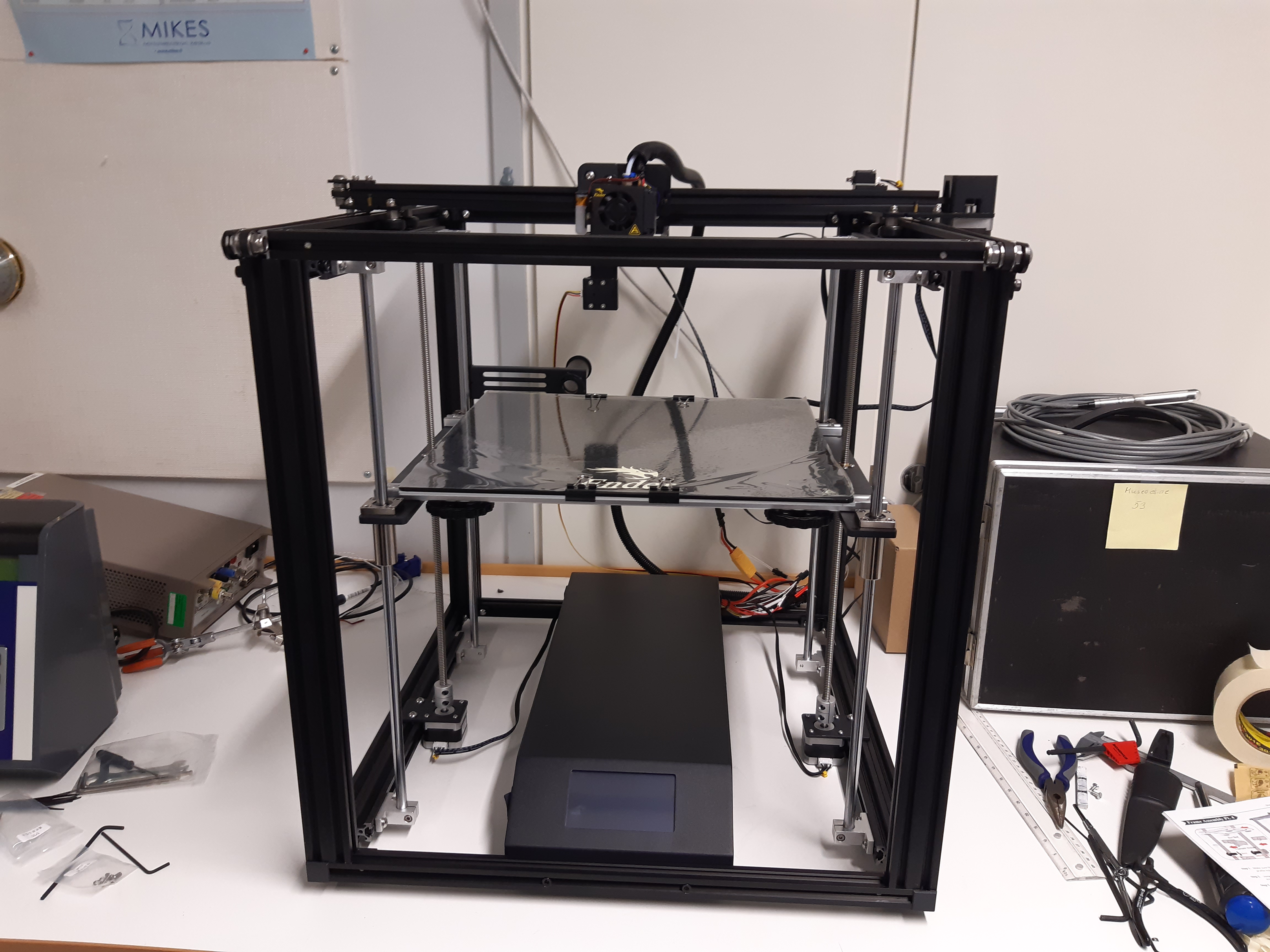}
        \subcaption{}
    \end{subfigure}
   \begin{subfigure}[t]{0.45\textwidth}
        \includegraphics[width=\textwidth, trim={20cm 0 30cm 20cm}, clip]{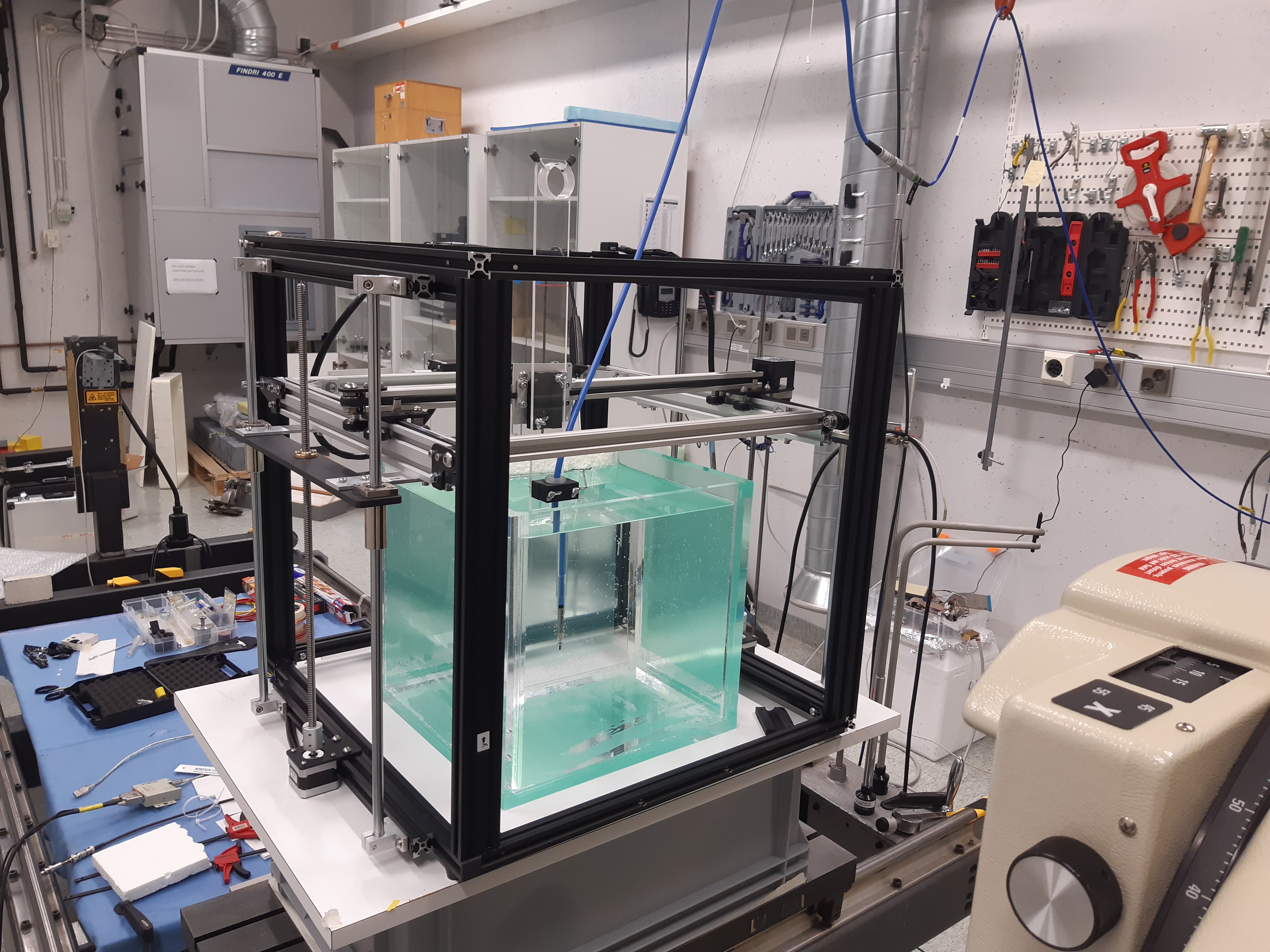}
        \subcaption{}
    \end{subfigure}
    \caption{a) Creality Ender 5 plus 3D-printer, b)~The DOSCAN scanner in an ionization chamber measurement setup, with a water phantom inside the scanner. The x-axis runs from left to right, y-axis from bottom to top, and the z-axis from front to back. In 3D-printing, the vertical axis is usually defined as the z-axis, but here the coordinate system was chosen so that the radiation beam in the laboratory is along the z-axis.}
    \label{DOSCAN}
\end{figure}

\section{Methods}

\subsection{Construction of the scanner}

The construction of the scanner began by modifying a Creality (Shenzhen, China) Ender 5 plus 3D-printer (figure \ref{DOSCAN}a). The coordinate system is defined in figure \ref{DOSCAN}. All movements are controlled by stepper-motors and the rotation of the motors is converted into linear motion with timing belts and pulleys (x- and z-axes) or with trapezoidal lead screws (y-axis). Because the scanner needed to move the detector in three dimensions, the printer was modified so that it could move a detector also on the vertical (y) axis. To allow this, the printer table was replaced with a chassis built with 20~mm~x~20~mm and 20~mm~x~40~mm V-slot aluminum profiles, the same profile that is used in the frame of the printer. The x-axis and z-axis parts were placed on the new chassis, and the whole x-axis - z-axis system was now moved up an down by the y-axis motors. The trolleys moved on the aluminum profile with V-slot wheels. The x-axis trolley was replaced with a self-made, more compact one. The original lead screws were used with the y-axis, but the timing belts for x- and z-axes were replaced. The y-axis motors were moved outside of the frame to increase the scan range of the x-axis. Our scanner used the stepper-motors from the printer.

The working principle of stepper-motors is to rotate a shaft by creating a magnetic field with coils. Bipolar stepper-motors have two sets of coils, and in the simplest mode of operation, full step mode, the currents through the coils are adjusted in four phases, or states. Changing from one state to the next one rotates the motor by one full step. Four steps in this mode of operation constitutes one sequence, after which the currents in the coils are in the same phase as in the beginning, and repeating the four phases keeps rotating the shaft. The currents can also be adjusted so that the magnetic field direction is between these full-step states, and this is called micro-stepping. The position of a detector can be calculated from the number of steps taken, and from the length of one step (step-size). 
The printer uses bipolar stepper-motors, with 200 steps per rotation of the shaft and a rated torque of 0.4~Nm. The nominal step-sizes of a full step for the scanner were 200~\um\ for the x- and z-axes and 20~\um\ for the y-axis.

The scan ranges were approximately 35~cm for the x- and y-, and 23~cm for the z-axis. The actual y-range is smaller in a radiation beam since the chassis should be sufficiently far away from the detector, preferably outside the beam to avoid excessive scattering. Placing a 30~cm high phantom inside the scanner frame would reduce the y-axis movement to 17.5~cm.

\subsubsection{Electronics} 

\begin{figure}
    \centering
    \includegraphics[width=0.5\textwidth, trim={0 5cm 0 0}, clip]{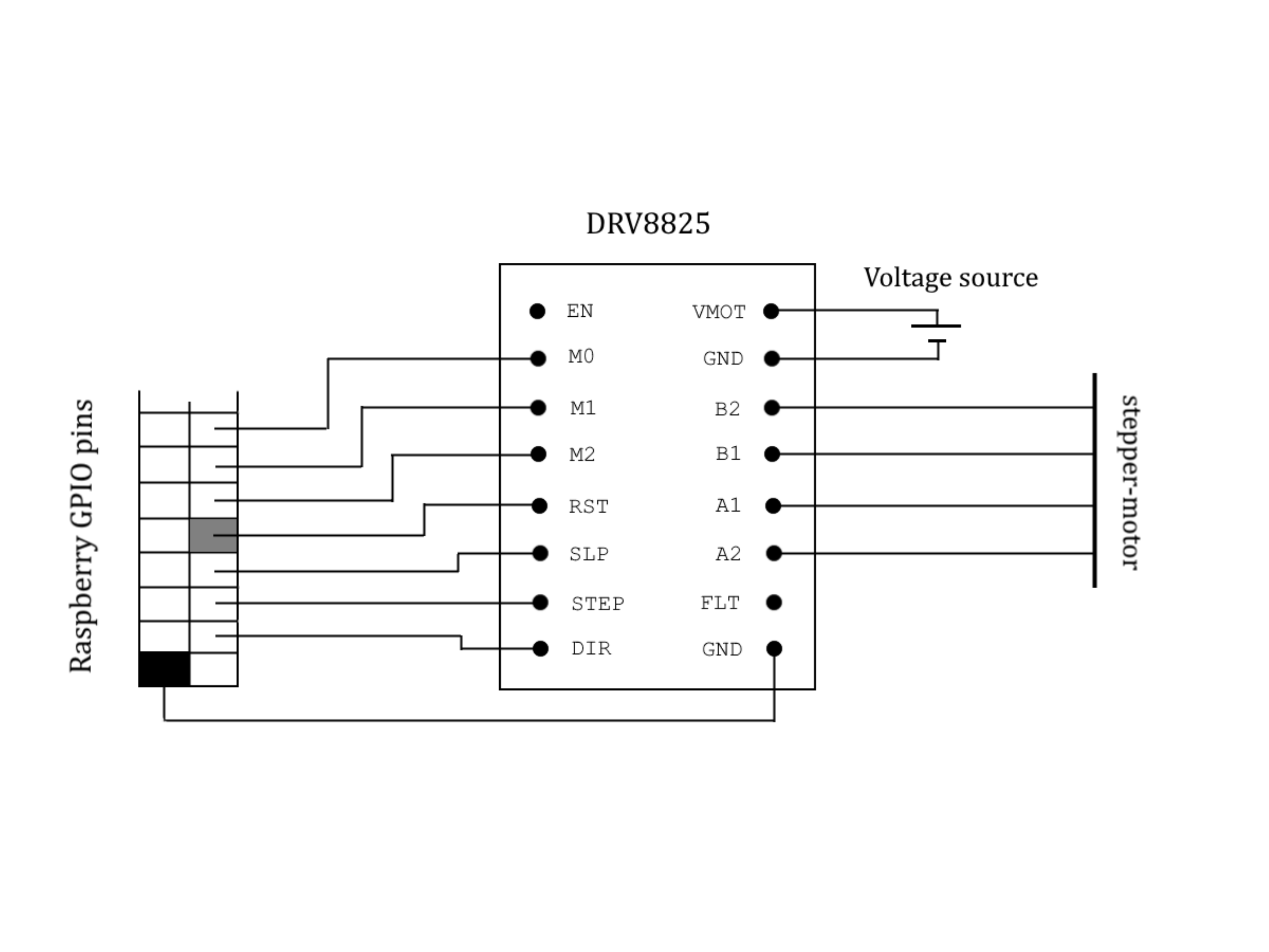}
    \caption{Illustration of the connections for the electronics for one stepper-motor. The Raspberry GPIO pins are not in realistic order. The gray pin represents a 3.3~V output and the black pin a ground pin.}
    \label{DRV}
\end{figure}

% schematic of electronics
To control the stepper-motors, we used a Raspberry Pi 4B (Raspberry Pi Foundation, Cambridge, UK) in combination with DRV8825 stepper-motor drivers (Texas Instruments, Dallas, USA). A DRV8825 adjusts the current in the coils of a motor, and hence can control the rotation of the motor shaft. The connections of a single motor are illustrated in figure \ref{DRV}. When a voltage is applied to the STEP pin of the driver (pin is pulled high), the motor takes one step. The direction is determined by pulling the DIR pin either high or low. By applying the micro-stepping pins M0, M1 and M2, the step-size can be adjusted to as low as 1/32 of a full step. In addition to improving the position resolution, micro-stepping makes the movement smoother. Pulling SLP (sleep) pin low shuts down the driver and cuts the current to the motor. The EN pin (enable) was left unconnected (high by default), and the RST (reset) pin was connected to a 3.3~V output of the Raspberry, meaning the pin is always high. A 45~V voltage source was attached to the drivers VMOT and ground pins to supply power to the motors. The connection to the motors was via JST PH 6-pin male connectors, where two of the six pins were unconnected.

Shutting the drivers for the duration of the measurement was necessary in some cases due to noise induced in a detector by the electronics. When the driver is shut down, the detector keeps still, but is more prone to movement by outside forces, and there might be small errors due to loosening tensions. To remedy these problems to some degree, the driver was kept on for a second before turning it off. Also, when the driver is shut down (SLP pulled low), it loses the memory of the state the motor was at, and when the driver is turned back on, it sets the motor to a default state. This state is defined here as 0, or full sequence. If the state was past the half sequence, the motor will turn forwards, and otherwise backwards. Turn to the default state could be compensated for in the software, but it still lead to an abrupt movement of 2 full steps at worst. Also, near half the sequence the direction to which a motor turned was not consistent, and the motors were prohibited to stop at such positions. 

The y-axis motors did not function reliably, and suffered sometimes form extra steps, or jerking motion. The problem was remedied somewhat by the small nominal step size of 20~\um\ per full step, but the reliability of the movement of the y-axis during a measurement could not be assured. All axes were always moved separately, so that only one motor was drawing power at a time.

\subsubsection{Software}

The scanner was operated with python code on the Raspberry, utilising the RPi.GPIO library \cite{RPiGPIO}. The package allows pulling GPIO pins on the Raspberry high and low, and hence the control of the motors. The loss of the memory of state when the motor is turned off was compensated for in the software by adding or subtracting a proper amount from the number of steps. The problem of the motor turning sometimes forward and sometimes backward when turned on near half the sequence was taken into account by not allowing the motor to be left in those states when turned off. The motor was turned forwards or backwards to the closest state where such behaviour was not observed. Therefore, in these cases the actual movement was not the requested one, but the position was known accurately.

The profile is scanned automatically in 1, 2 or 3 dimensions with measurement intervals in each direction provided by the user. The program can be told to wait for a certain time after moving while the detector measures, or it can communicate with the software used with a detector and continue the scan when the measurement is completed. In the latter case, a socket connection, for example, can be established between a PC and Raspberry Pi. Continuous movement is also possible. 

The scanner was named DOSCAN, a wordplay with the abbreviation DOS of the laboratory name.

\subsection{Verification of position accuracy}

The position accuracy and repeatability of the scanner was studied in three ways: by using a caliper, a laser displacement sensor, and by making repeated measurements of a $^{60}$Co irradiator beam profile. The caliper could be read with 10~\um\ resolution, although the measurement uncertainty was estimated to be larger. Due to limited range of the caliper, the measurements were taken in two parts for the x-, and three parts for the z-axis, using different points from which to measure the distance to the x-axis trolley (called here reference points). For x-axis, the accuracy could be determined throughout the measurement range by measuring the distance between the reference points, but with increased uncertainty for the second set of measurements. For the z-axis, the distance between the reference points could not be accurately measured, and the accuracy of the movement could be determined only within each of the three sets.

In the measurement with a Micro-Epsilon (Ortenburg, Germany) ILD1700 laser displacement sensor, a 30~x~30~x~30~cm$^3$ water phantom was put inside the scanner frame, an IBA FC65-G ionization chamber (IBA Dosimetry GmbH, Schwarzenbruck, Germany) was attached to the trolley, and the distance was measured to the closest point on the chamber surface. The phantom wall material was PMMA, and the thickness of front wall 1~cm. A thin tape was put on the chamber to obtain a smoother surface. The laser was attached outside of the phantom front wall. The distance measurement with the laser is based on triangulation of the scattered light from the measured object, and in our setup, the refraction of light at the air-phantom wall-water interfaces affects the reading of the laser. However, from calculation of the angle of light coming to the sensor as a function of distance in the phantom, the reading of the laser changes practically linearly with distance also in the phantom within the approximately 2~cm measurement range. The laser reading could have been calibrated to show depth in the phantom (distance from the outer surface of the front wall), but the calibration points cover only a small part of the measurement range around 5~cm depth, and the results at the edges of the range were clearly erroneous. Therefore only the linearity and repeatability could be tested with the laser.

Dose-to-water profiles, or more accurately ionization chamber current profiles for the $^{60}$Co irradiator (GBX200, Best Theratronics Ltd., Ottawa, Canada) were measured with a PTW (Freiburg, Germany) 31015 Pinpoint chamber and a Keithley 6517B electrometer (Tektronix, Beaverton, USA). The profile measurements were along the x-axis. The chamber was attached to the trolley, and positioned at 5~cm depth in the phantom with the IDL1700 laser. The distance of the phantom front wall from the $^{60}$Co source was 95~cm and the field size at the 5~cm depth was 10~cm~x~10~cm, as recommended in \cite{TRS398}. The profile was scanned in either 0.5~cm or 1~cm intervals from -8~cm to 8~cm from the center of the beam. The chamber tip was pointing downwards. The detector stayed stationary while the current was measured at each position and was moved back to the starting position after the scan. The scan was repeated five times. The profile was also measured in 1~mm intervals for visualisation of the beam shape. 
In all measurements, the motors were turned off after moving, and the step-mode was 1/4 steps.

\subsection{Volume averaging correction}

The GBX200 dose-profile was also measured using the DOSCAN scanner and a PTW 31010 Semiflex chamber, to investigate the effects of the chamber size on the profile measurement. The Semiflex chamber cavity length is 6.5~mm and radius 2.75~mm, whereas the length of the Pinpoint chamber cavity is 5~mm and radius 1.45~mm. The purpose of the volume averaging correction is to take into account the averaging of the dose in the chamber cavity volume, and correct the dose to the value at the center of the cavity volume. The volume averaging correction factor for the chamber at position $(x_0,y_0,z_0)$ can be written as
\begin{equation}
    k_\text{vol}(x_0,y_0,z_0) = \frac{\int_V \text{d}x\text{d}y\text{d}z}{\int_V \frac{D(x,y,z)}{D(x_0,y_0,z_0)}\text{d}x\text{d}y\text{d}z},
    \label{kvolorg}
\end{equation}
where $D(x,y,z)$ is the dose profile. The integration is done over the cavity volume, and hence the numerator in \eqref{kvolorg} is the cavity volume. In \cite{TRS483}, the volume averaging correction is calculated along the length of the chamber (y-axis) at the center of the beam. Here, since most of the interest is at the values at the edge of the beam and the most of the non-uniformity in beam shape can be assumed to be in x-direction, the volume averaging correction is calculated along the x-axis, and y- and z- dose distributions are assumed to have a minor effect. Then, for the cavity volume center at position $x_0$ on the x-axis, equation \eqref{kvolorg} can be written as a one dimensional integral
\begin{equation}
    k_\text{vol}(x_0) = \frac{\pi r^2}{\int_{x_0-r}^{x_0+r} \frac{D(x)w(x_0-x)}{D(x_0)}\text{d}x},
    \label{kvol}
\end{equation}
where $r$ is the chamber cavity radius and $w(x_0-x)$ the cavity thickness on the xz-plane in z-direction distance $|x_0-x|$ away from the center of the cavity. Since the detector tip is pointing downwards, the cavity cross-section on the xz-plane is circular, and $w(x-x_0)=2\sqrt{r^2 - (x-x_0)^2}$. The $w$ parameter takes into account that more signal is generated along the x-axis at thicker parts of the cavity. The profile shape along the y- and z-axes also affect the volume averaging correction, but the effect was not investigated.

When the profile is measured with a chamber for which the volume averaging needs to be taken into account, the problem is that $k_\text{vol}$ has to be calculated from the dose-averaged measured profile. The process of calculating $k_\text{vol}$ can be iterated, meaning that new $k_\text{vol}$ factors are calculated from the once $k_\text{vol}$ corrected profile, and the profile should approach the actual shape when the process is repeated. However, the iteration process may amplify the scatter of the profile.

The volume correction was calculated for the beam-profile measured with the Semiflex chamber. The chamber was moved in 1~mm intervals at the 5~cm depth. The dose-profile $D(x)$ was obtained by making a 3rd order polynomial fit to the measured values $\pm$4~mm away from the cavity volume center at each measurement point, and the integral in equation \eqref{kvol} was calculated numerically in 10~\um\ intervals.

\section{Results and discussion}

\begin{figure}[b!]
    \centering
    \includegraphics[width=0.45\textwidth]{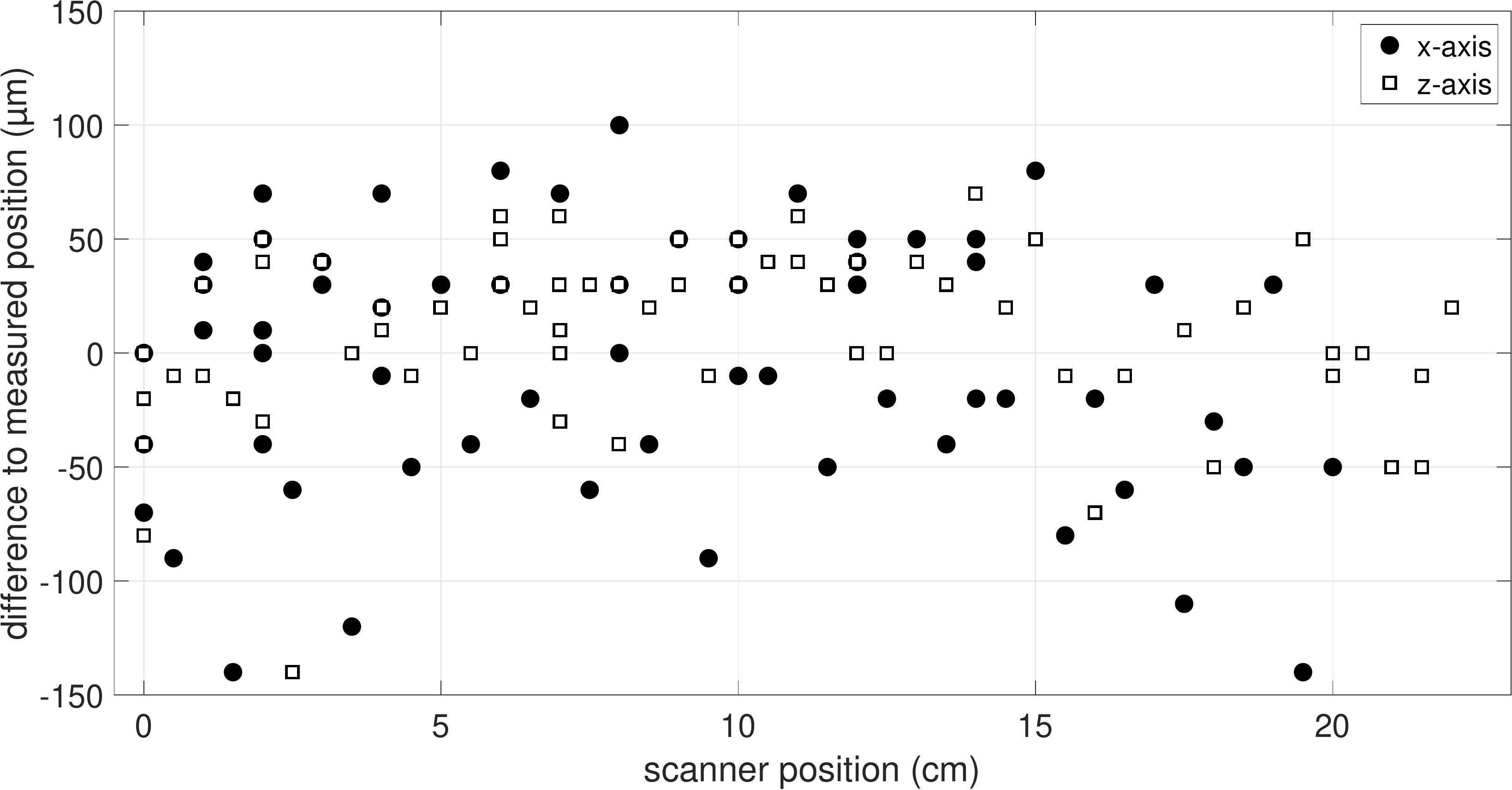}
    \caption{Difference between position calculated with scanner steps and the position measured with a caliper. The values for both axes constitute of multiple sets of measurements, with a separate reference point (point where the difference is defined to be zero) for each set. For x-axis, the sets were at distances 0~cm - 13~cm and 14~cm - 20~cm, and for the z-axis at 0~cm - 7~cm, 7~cm - 14~cm and 15.5~cm - 22~cm. The starting point (0~cm) was arbitrary. The detector was moved back and forth, and hence there can be multiple measurement points at the same distance.}
    \label{caliper}
\end{figure}

The measurements with the caliper showed that the length of one full step was very precisely 0.2~mm for the x- and z-axes, since the deviation between scanner position calculated from the number of steps (using the 0.2~mm step-size) and the measured position did not grow systematically with distance (figure \ref{caliper}). The results show an accuracy better than $\pm$150~\um\ over the scan range. However, the measurements had to be done in multiple parts using different points from which to measure the distance with the caliper, and the accuracy could be measured with increased uncertainty for the second set of measurements for the x-axis, and only within each measurement set for the z-axis. Still, the results do not indicate the error in position to grow with the distance.

\begin{figure*}[htb!]
    \centering
    \begin{subfigure}[t]{0.45\textwidth}
        \centering
        \includegraphics[width=\textwidth]{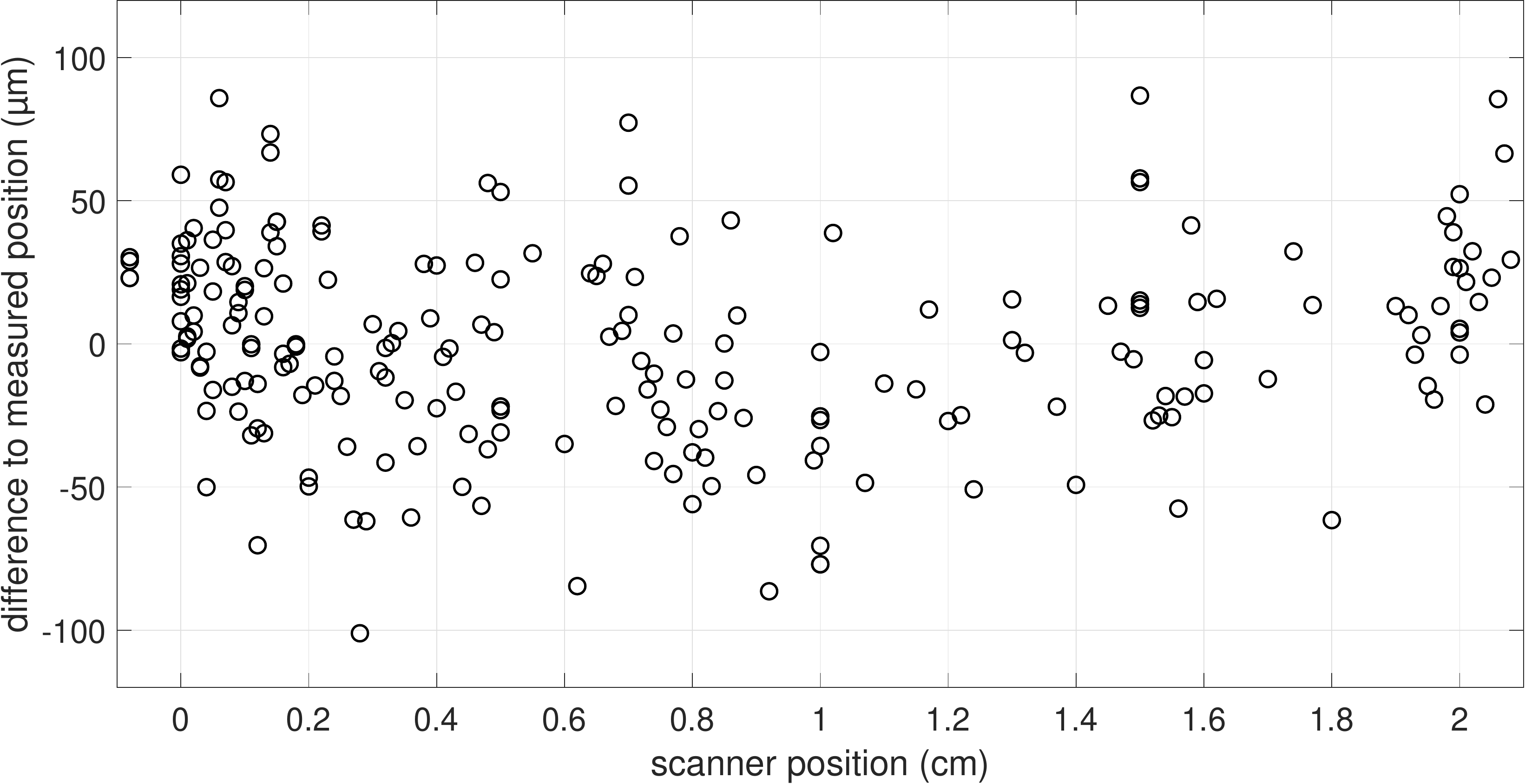}
        \caption{}
    \end{subfigure}
    \hspace{1cm}
    \begin{subfigure}[t]{0.45\textwidth}
        \centering
        \includegraphics[width=\textwidth]{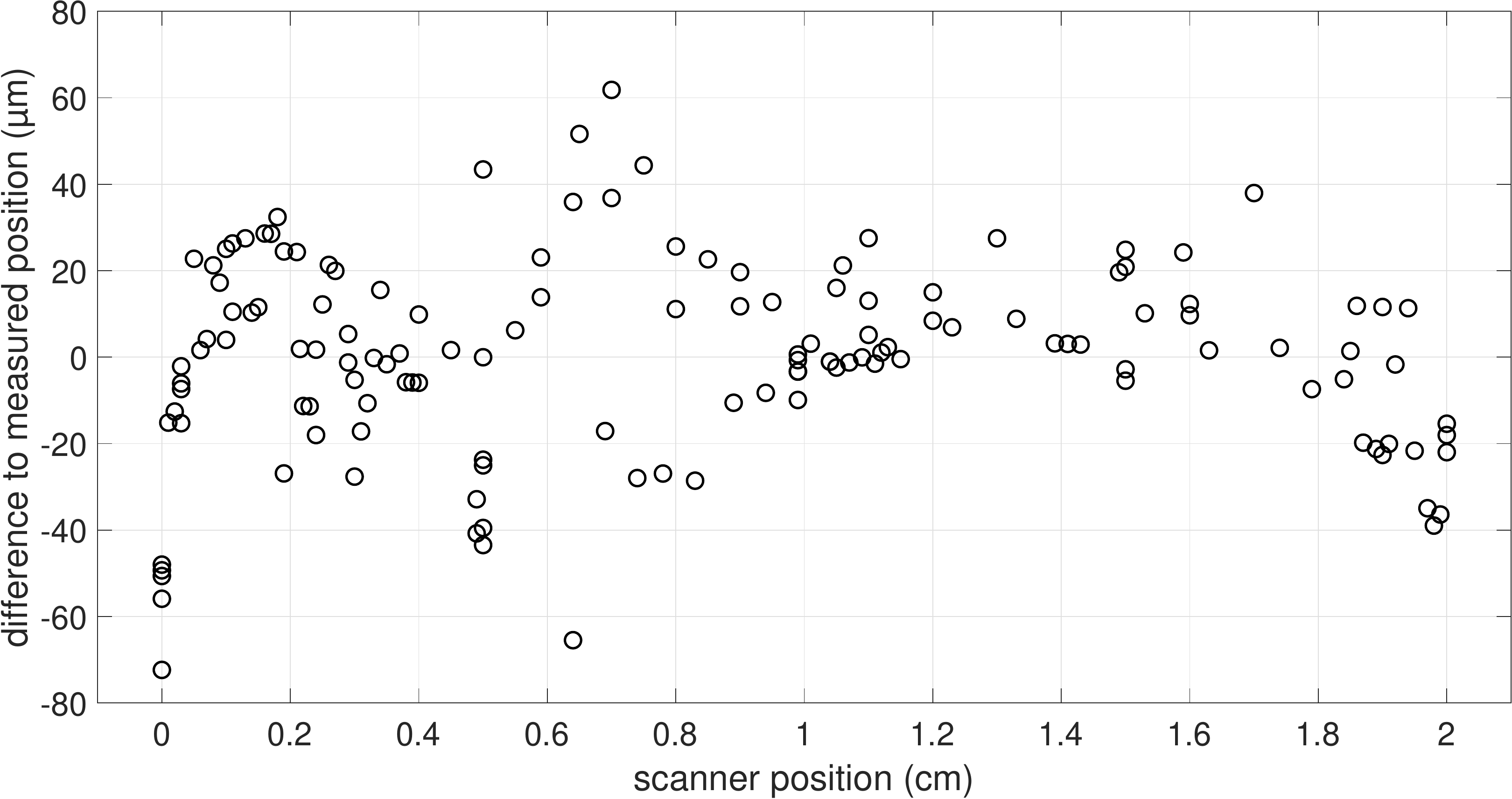}
        \caption{}
    \end{subfigure}
    \begin{subfigure}[c]{0.45\textwidth}
        \centering
        \includegraphics[width=\textwidth]{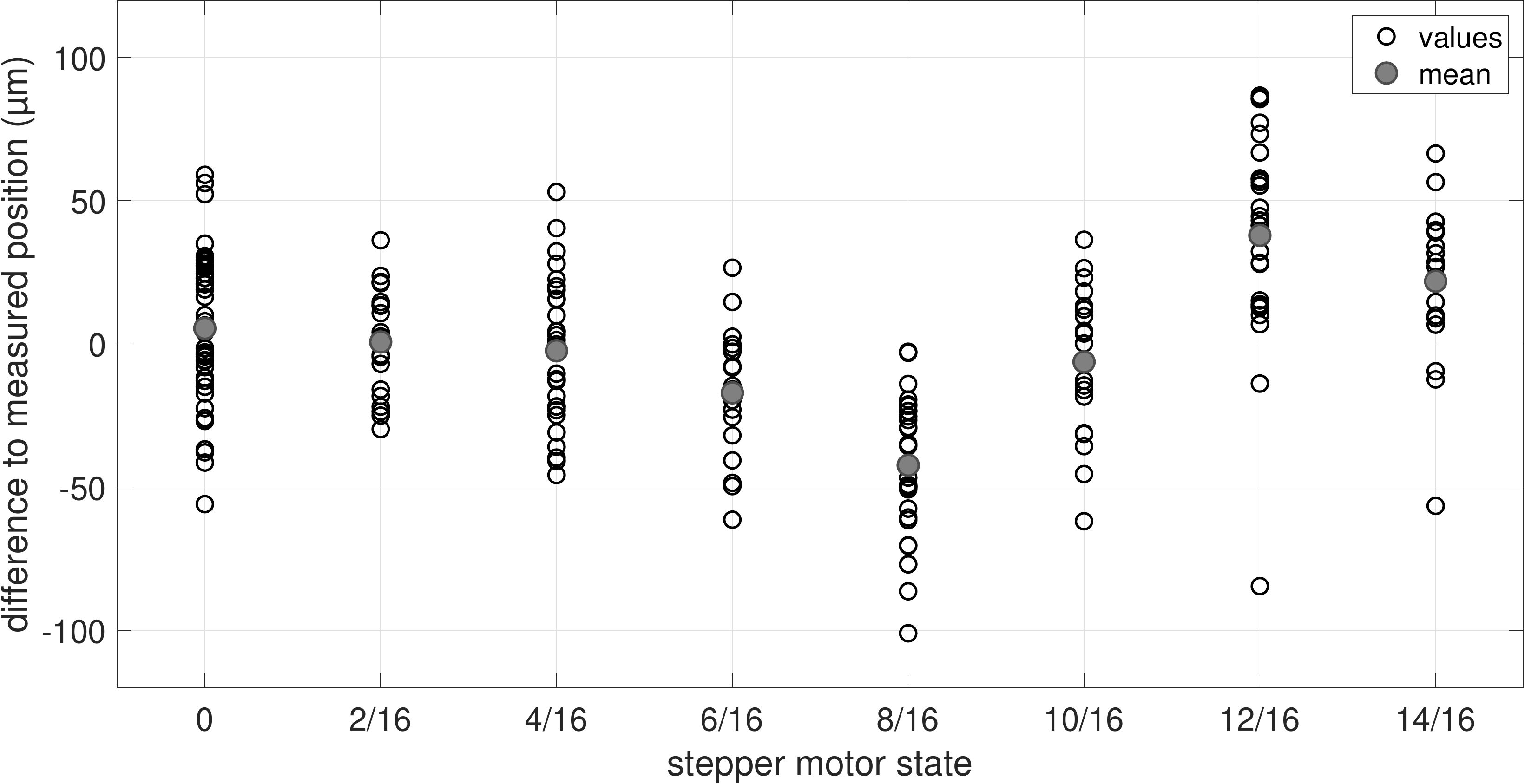}
        \caption{}
    \end{subfigure}
    \hspace{1cm}
    \begin{subfigure}[c]{0.45\textwidth}
        \centering
        \includegraphics[width=\textwidth]{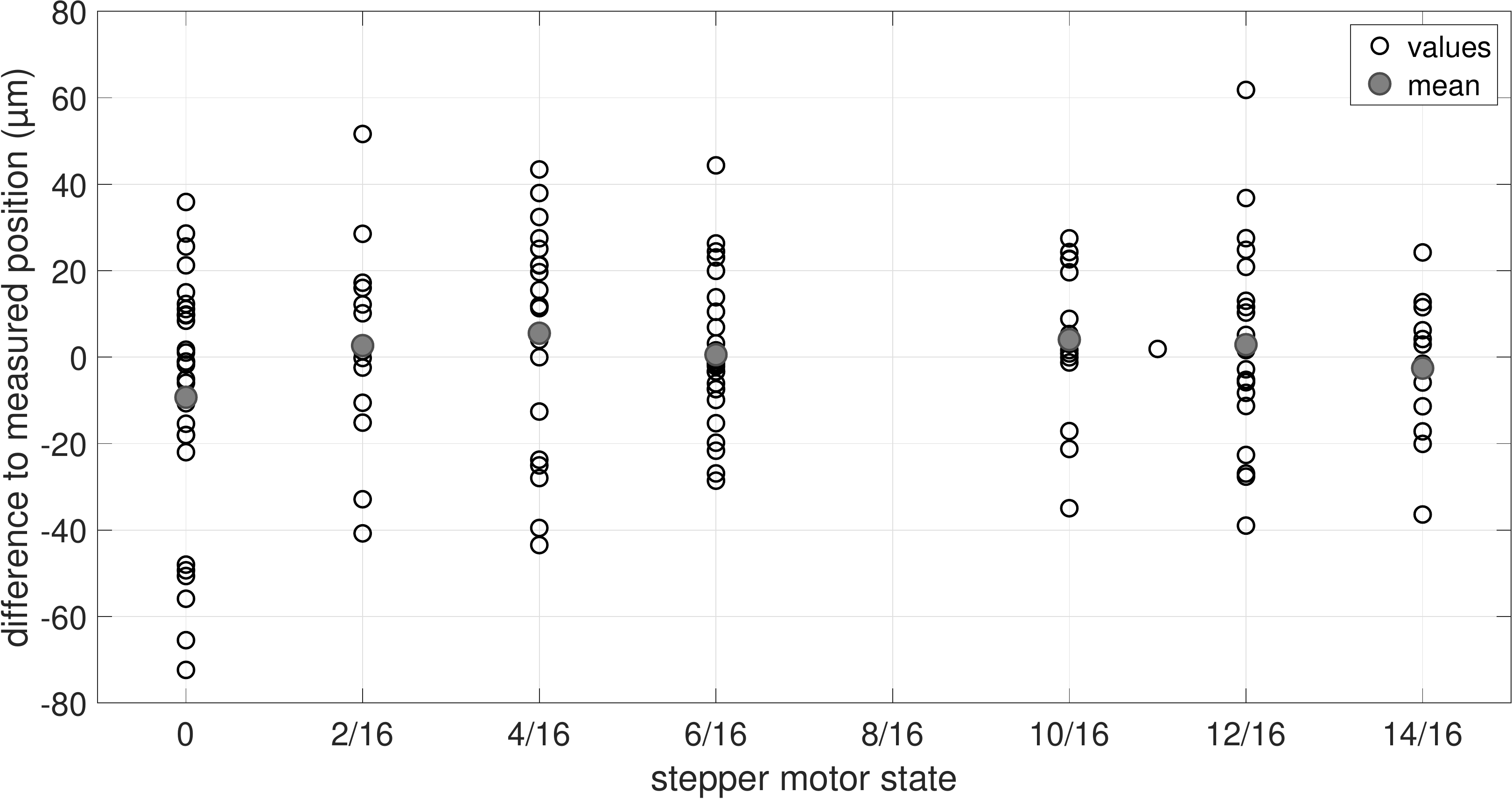}
        \caption{}
    \end{subfigure}
     \begin{subfigure}[b]{0.4\textwidth}
        \centering
        \includegraphics[width=\textwidth]{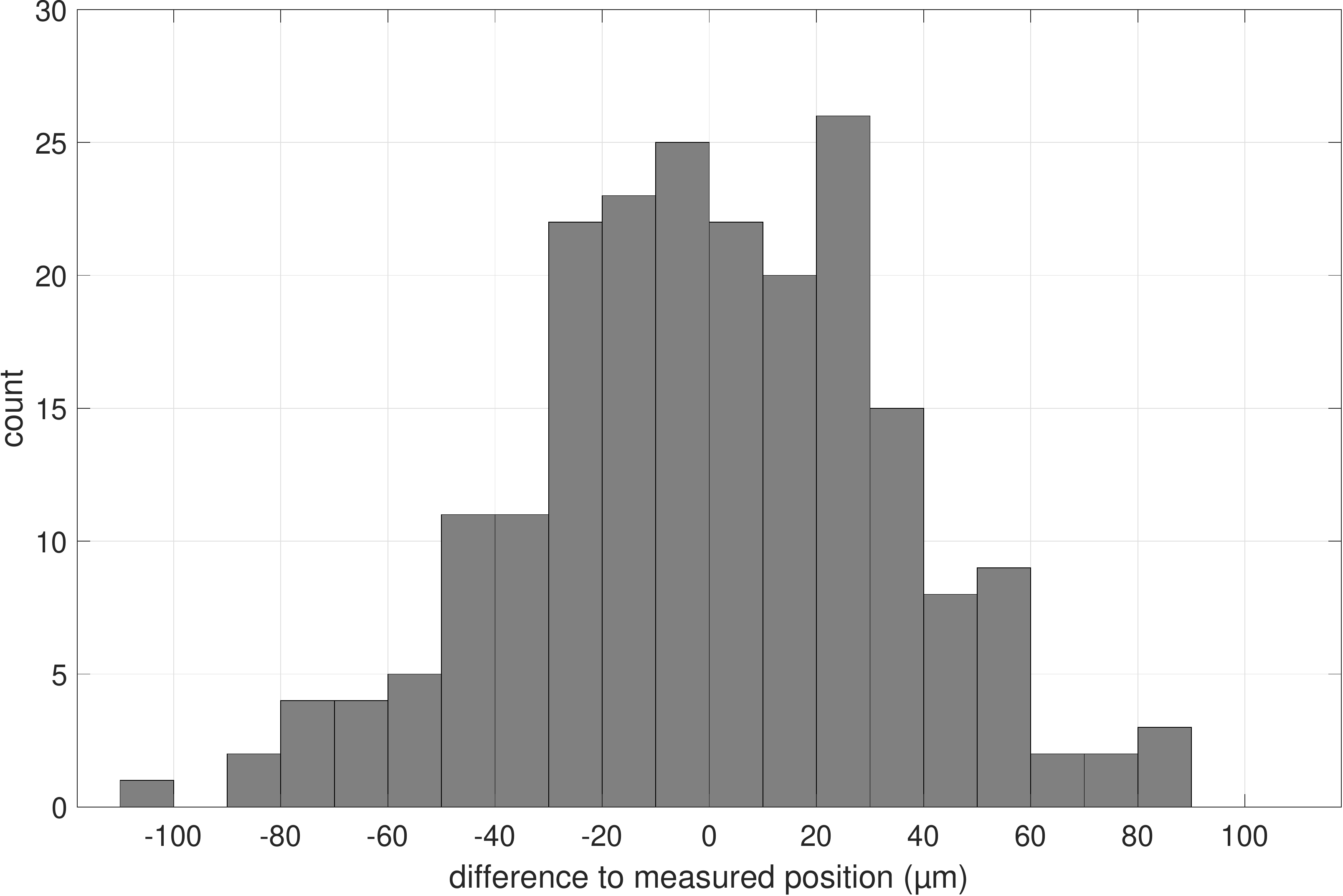}
        \caption{}
    \end{subfigure}
    \hspace{1.5cm}
    \begin{subfigure}[b]{0.4\textwidth}
        \centering
        \includegraphics[width=\textwidth]{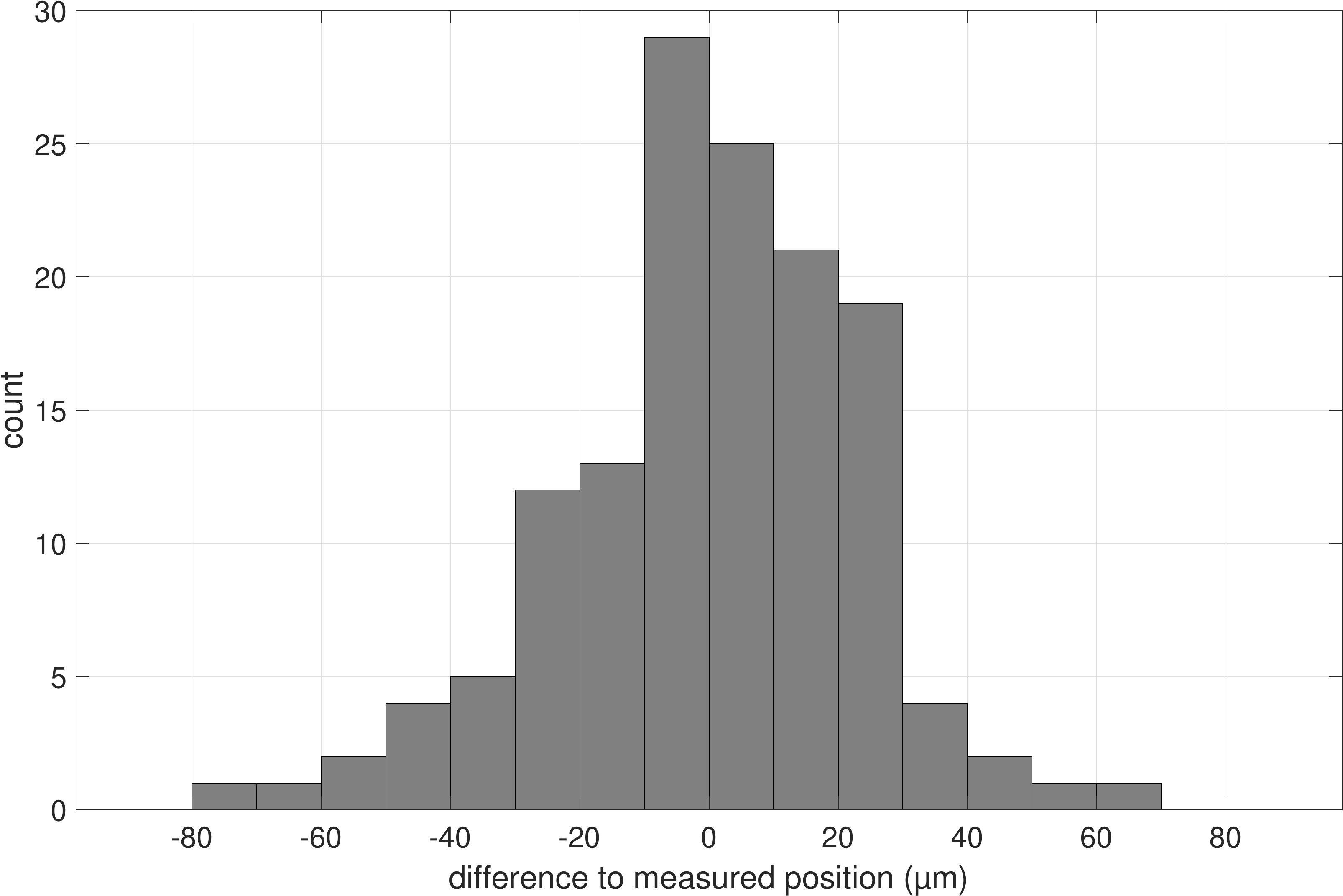}
        \caption{}
    \end{subfigure}
    \caption{a) and b) Difference between position calculated from the scanner steps and the position measured with the IDL1700 laser displacement sensor as a function of distance for the x-axis. c) and d) The difference as a function of the stepper-motor state. The mean of the difference was calculated for each state and is plotted as gray circles. e) and f) Distribution of the difference calculated with a 10~\um\ bin size. a), c) and e) show the results for the x-axis, and b), d) and f) for the z-axis. For the z-axis, an outlier at 1.93~cm with 147~\um\ difference was omitted from the plots.}
    \label{laser}
\end{figure*}

The results for the difference between the IDL1700 laser displacement measurement sensor and the calculated position, as a function of distance and stepper-motor state, are in figure \ref{laser}. Since the sensor could not be calibrated to give the absolute position throughout its measurement range, the measured position was obtained from a linear fit to the scanner position - laser reading data. In other words, the scanner was assumed to be accurate on average, and comparison of the calculated scanner position and the laser measurement shows the linearity and repeatability of the movement. The assumption of the scanner being accurate on average is supported by the caliper measurements.

The deviation from the linear fit did not change significantly as a function of distance for either axis. For the x-axis, the state of the motor had a significant effect on the position for the states 8/16 and 12/16 (figure \ref{laser}c). This would be explained by a higher magnitude of negative current in one set of the coils, but the effect was not investigated further. The distribution of the difference (figures \ref{laser}e and \ref{laser}f) was wider for the x-axis (standard deviation \SI{35}{\micro m}) than for the z-axis (standard deviation \SI{26}{\micro m}). Excluding the states 8/16 and 12/16 from the x-axis data would reduce the standard deviation to 27~\um. With the exception of one measurement for each axis, the measured and calculated positions were within 100~\um\ from each other. The uncertainty of the laser measurement was not estimated, and how much of the variance in the results is caused by the scanner and by the laser is not known.

\begin{figure*}[t!]
    \centering
    \begin{subfigure}[c]{\textwidth}
        \centering
        \includegraphics[width=0.5\textwidth]{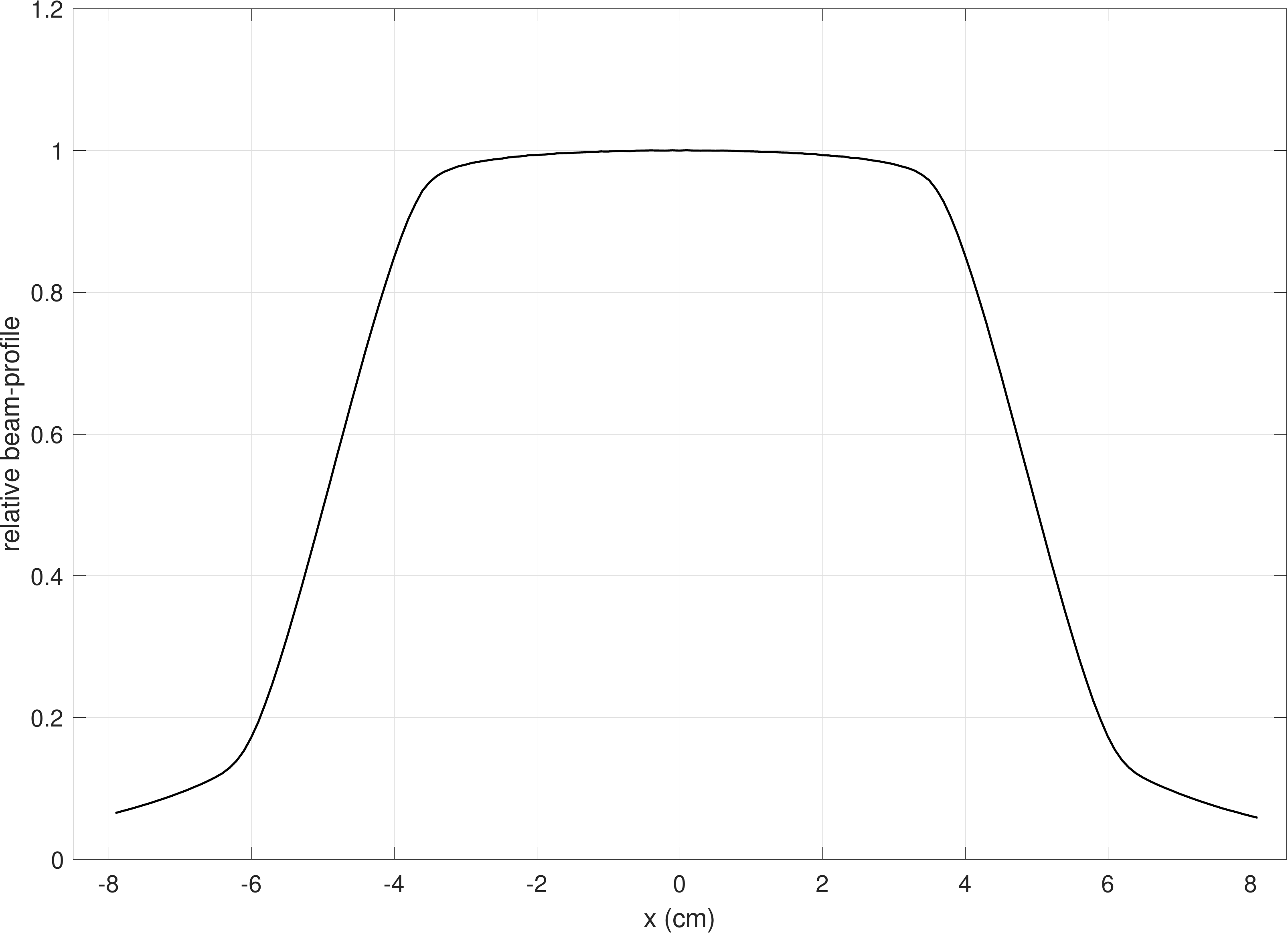}
        \caption{}
    \end{subfigure}
    \begin{subfigure}[c]{\textwidth}
        \centering
        \includegraphics[width=0.5\textwidth]{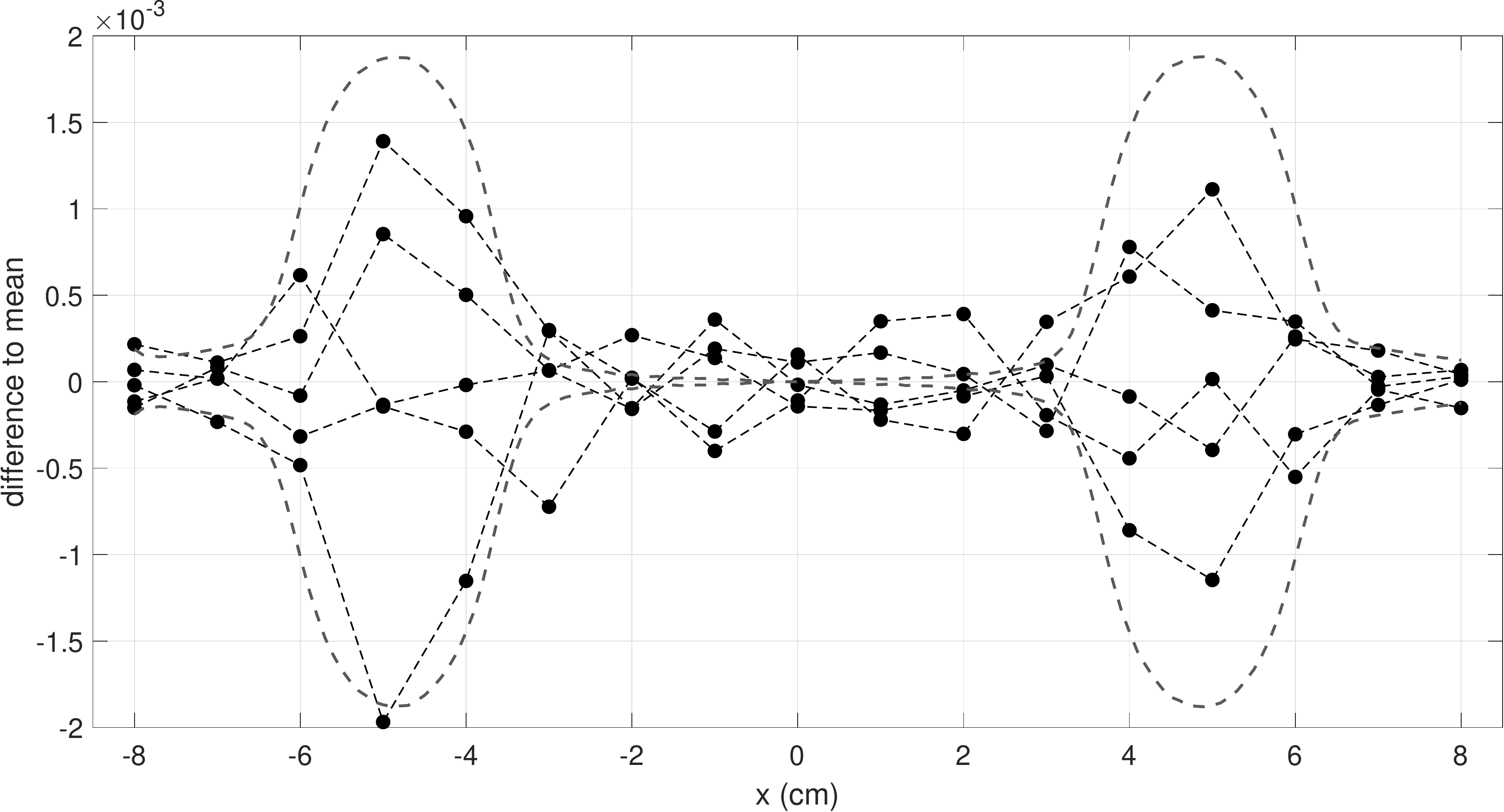}
        \caption{}
    \end{subfigure}
    \caption{a) Relative GBX200 dose-profile at 5~cm depth in the water phantom at 100~cm distance from the source. b)~Difference between the five repeated scans (dashed line and open circles), where the value being compared to is the mean of the results at each position. The data is normalized by maximum current of all the scans to convert the current values to relative profile. The dashed gray line is the estimated error that would be caused by $\pm 50$~\si{\micro\meter} shift in the chamber position, calculated from the gradient of the profile.}
    \label{beam_meas}
\end{figure*}

The results from repeated beam-profile measurements are in figure \ref{beam_meas}b. The results from the more detailed scan are also shown to visualize the profile shape (figure \ref{beam_meas}a). At the center of the beam, the repeated measurements do not give relevant information about the repeatability of the scanner, since quite large changes in position along the x-axis would have only a minor effect on the measured current. Here the deviations from the mean are due to variance of the current measurement, and possible small differences in the depth of the chamber. At the edges of the beam, the gradient of the profile is steep and small changes in the detector position would be detected through a change in the current. This is seen in figure \ref{beam_meas}b as a higher variation at positions of larger gradient. The differences in the measured current could be explained by a maximum deviation in the position of slightly more than 50~\si{\micro\meter} from the mean (the dashed gray curve in figure \ref{beam_meas}b). Two of the scans were done with 0.5~cm intervals instead of 1~cm to test whether smaller interval would affect the results (only the data at the 1~cm interval positions is shown), but such an effect was not detected.

The vertical y-axis suffered from unknown accuracy problems, and although it did function as intended during some measurements, we do not trust the positioning to be reliable. With better electronics, a high accuracy should be possible also for this axis because of the small step size of 20~\si{\micro\meter} per one full step of the stepper-motor. Other solutions than using the Raspberry and the DRV8825 drivers, for example the electronics from the printer itself, are under investigation.

For PTW MP3-XS, IBA Blue Phantom$^2$, Standard Imaging Dose View 3D (Middleton, USA) and Sun Nuclear 3D Scanner (Melbourne, USA), the manufacturers state the position accuracy to be $\pm0.1$~mm \cite{MP3, BluePhantom, Doseview3D, SN_3D_scanner}. Similar values have been given in \cite{Saenz2016}. Therefore, the accuracy of the DOSCAN is close to that of commercial products. Also, the position resolution for the DOSCAN is better than the suggested value of 0.1~mm from \cite{TRS483}.

\begin{figure*}[t!]
    \centering
    \includegraphics[width=0.6\textwidth]{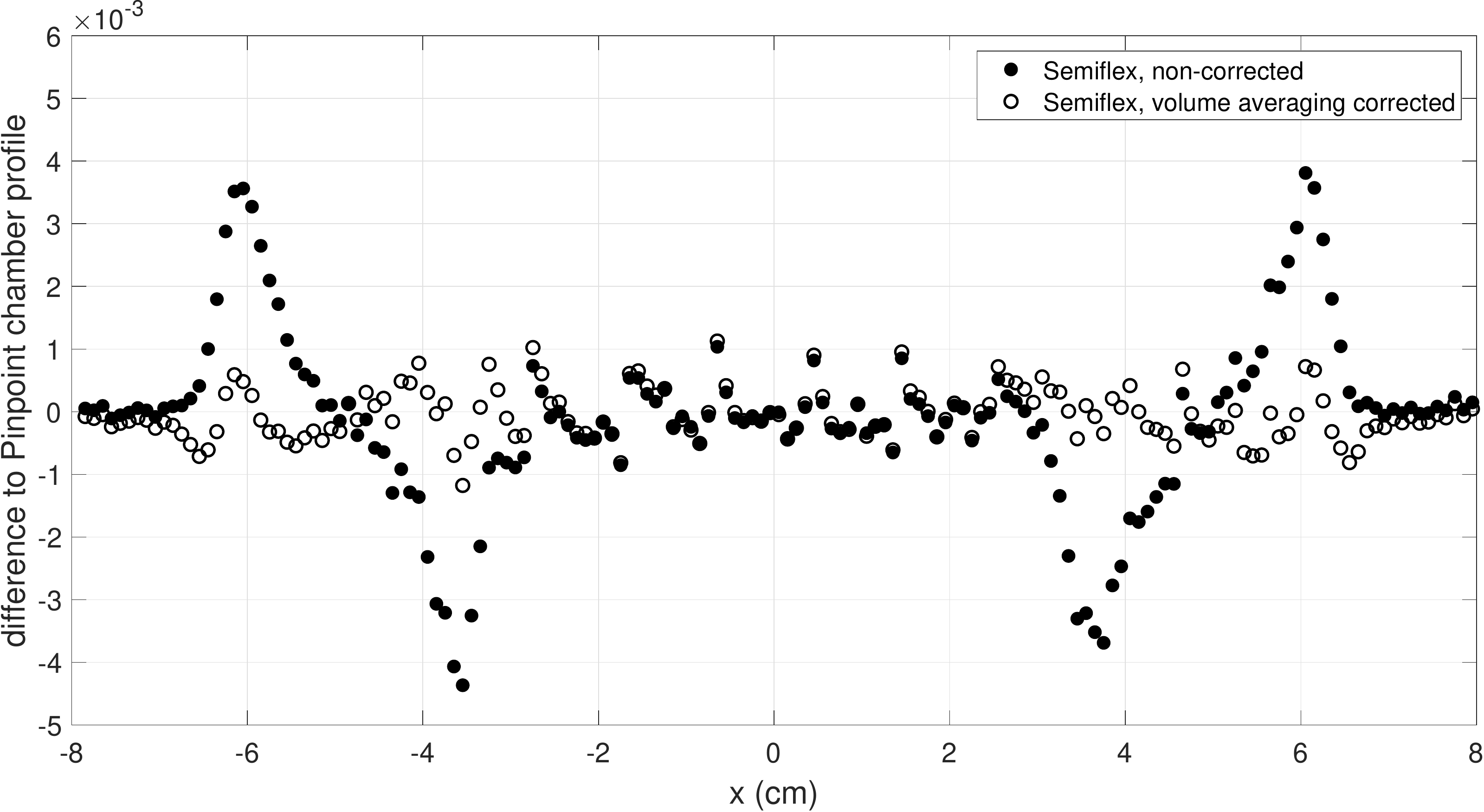}
    \caption{Comparison of the measured beam-profiles of the PTW Pinpoint 31015 and Semiflex 31010, with and without applying volume averaging correction factors calculated with \eqref{kvol}. The y-axis shows the absolute difference of the (relative) profiles.}
    \label{Semiflex_vs_pinpoint}
\end{figure*}

The difference of beam-profiles measured with the PTW 31015 Pinpoint and Semiflex 31010 chambers is visualized in figure \ref{Semiflex_vs_pinpoint}. The x-axis values were shifted so that zero position was exactly between the half-values of the profile, and this allowed the comparison of the profiles without the uncertainty of detector positioning. As expected, the beam corners are smoother with the larger Semiflex chamber due to dose-averaging in the cavity volume, and the differences are highest where the absolute value of the second derivative of the profile is largest. After calculating the $k_\text{vol}$ factor with \eqref{kvol} at each measurement point for the Semiflex chamber, the Semiflex and Pinpoint profiles are identical almost within 0.001 (absolute difference of the relative dose-profiles). The calculation of the $k_\text{vol}$ factors was not iterated, since the differences after the second iteration were insignificant, and the scatter was amplified. The $k_\text{vol}$ factors were also calculated for the Pinpoint chamber, but were not applied, since the correction was below 0.001.

\section{Conclusions}

The DOSCAN 3D scanner was built from relatively cheap and simple components and showed, within the range of the distance measurement equipment, a typical position accuracy better than 100~\um, and maximum deviations of approximately 150~\um. Results from the position measurements, and measurements with the ionization chamber show that the scanner is sufficiently accurate for dose-profile measurements.

Using a PTW 31010 Semiflex chamber for the GBX200 $^{60}$Co irradiator beam-profile measurement lead to differences of almost 0.005 in the relative profile compared to the profile measured with a PTW 31015 Pinpoint chamber. However, when volume averaging correction factors calculated along the measurement axis were applied to the results measured with the Semiflex chamber, the maximum difference decreased to approximately 0.001.

Building a similar scanner as DOSCAN from scratch might be a valid solution, and would allow more freedom in the range of the scans. The scanner as it is uses mostly parts available in 3D-printer stores, and hence a similar device could be built from components bought separately from a complete 3D printer, perhaps with little machining. Building a device such as DOSCAN for purposes of moving non-heavy equipment with high accuracy can be a valid option for laboratories, radiation related or not, compared to commercial products.

\section*{Acknowledgements}

This study was performed in the framework of the Academy of Finland project, number 314473, \textit{Multispectral photon-counting for medical imaging and beam characterization}. Support for this work was also received from the Helsinki Institute of Physics.

\bibliographystyle{unsrt}
\bibliography{bibliography_scanner_paper}

\end{document}